\documentclass[aps,twocolumn,amssymb,showpacs,amsmath,preprintnumbers]{revtex4}
\usepackage{graphicx}
\usepackage{bm}

\begin{document}

\title{Analytical evidence for quantum states in aqueous vanadium pentoxide with positron lifetime spectroscopy}
\author{L V Elnikova}

\address{A. I. Alikhanov Institute for Theoretical and Experimental Physics, \\B. Cheremushkinskaya st. 25, Moscow 117218, Russia}

\begin{abstract}
The possibility of registration of quantum states, such as
the coalescence of droplets (tactoids) in the sol phase of aqueous
vanadium pentoxide V$_2$O$_5$, with positron annihilation
lifetime spectroscopy is discussed. The decrease of the long-living positronium (Ps) lifetime term in the result of the coalescence of V$_2$O$_5$ tactoids is predicted.
\end{abstract}
\maketitle
\section{Introduction}
The inorganic liquid
crystal (LC) state of V$_2$O$_5$ was discovered by Z\"{o}cher \cite{Zocher}
in 1925, the detailed historical review of the studies
of mesophases in V$_2$O$_5$--H$_2$O system was found in \cite{Sonin}. A number of experimental investigations was devoted to clarification of structure and physicochemical properties of aqueous V$_2$O$_5$ mesophases with optical, X-ray and NMR spectroscopy \cite{Sonin}, \cite{Kazna}, \cite{PD}.

This knowledge acquires more relevance,
as far as areas of the realization of vanadium oxides are expanding with
the synthesis of nanostructures, the manufacturing of cathodes in lithium batteries or counter-electrodes in electrochromic devices, antistatic coating in the photographic industry, conducting matrix in glucose biosensors \cite{PD}, \cite{Livage} and so on.

Because of anisotropic (spindle-like) form of droplets (tactoids, how Z\"{o}cher named them), aqueous $V_2O_5$ sols form nematic phase \cite{Kazna}, \cite{PD}.
Sizes of a tactoid particle depend on ageing time due to the type of ageing processes due to which they are carrying out (e.g. ageing in the Biltz sol, or in the M\"{u}ller sol, either in the Prandtl and Hess sol). For example, their average length was found to be 0.4 $\mu$m after one year of ageing \cite{Sonin}. A reversible sol-gel transition is observed
for a vanadium concentration corresponding to
V$_2$O$_5$,250H$_2$O (where number 250 means $n$ in formula V$_2$O$_5$,$n$H$_2$O; [V]$\approx$ 0.2 mol/l=18 mass $\%$) \cite{Sonin}, \cite{PD}. The sol phase lies approximately between 250 and 400 $n$, and $n$=400-600 corresponds to the biphasic region, whereas the region of $n>$600 has occupied by isotropic sols at the phase diagram of V$_2$O$_5$,$n$H$_2$O \cite{Pryanik}.

The attractiveness of PAS methods to study the droplet coalescence in V$_2$O$_5$--H$_2$O sols is motivated by significant troubles of their direct
registration with optical spectroscopy and X-ray diffraction.

In lifetime PAS of the aqueous V$_2$O$_5$ sols, both water and tactoid collisions with the positron beam are measurable.
The Ps formation in water is more or less interpreted (\cite{Mogensen}). Whereas the sol phase of vanadium pentoxide is puzzle in this meaning.

There is a stimulated example of the hydration of V$_2$O$_5$ powders in CaO-ZrO$_2$-SiO$_2$ glasses, where lifetime positron annihilation data revealed changes in magnetic properties of the whole system caused by influence of V$^{5+}$, V$^{4+}$,
and V$^{3+}$ paramagnetic ions \cite{egypt} represented in
the same mole fraction as that in aqueous sols.

The basic data, received or with the Doppler broadening, either with angular correlation technique \cite{Johnson}, exhibited the role of $d$-electrons of transition metals (including vanadium), their average value of the annihilation rate $\lambda_e$ (for any transition metals) is taken to be 1.5 ns$^{-1}$.

Also there are developed the theoretical modeling for predictions of positronium spectra, which are based
on (\cite{Puska} and references therein) calculations of electron
states concerning to the measuring variation characteristics of Ps,
carrying out due to \textit{ab initio} or Molecular Dynamics (MD) principles. However, important analytical methods for iono-covalent oxides \cite{Lenglet} contain in the models which include the
electronegativity and the acidity scales supplied with the classification of partial negative charges \cite{Zhang}.

Positron annihilation studies of
V$_2$O$_5$--H$_2$O sols, especially during a regime of the coalescence of
droplets, may helpfully specify their physicochemical and structure properties.

\section{Mechanism of Ps formation in V$_2$O$_5$--H$_2$O sols}
Analyzing the facts on condensation processes at the chemical
synthesis of vanadium pentoxide polymers (see [21] in \cite{PD}),
and on an orientation
of tactoid particles (Figure 1)
in a magnetic field, one may suppose, that they
bear some negative charge, as well as single V$_2$O$_5$ molecules \cite{PD}, $\approx$0.2e$^-$ per V$_2$O$_5$; in the first case, a charge
arises from the acid dissociation of V$^{5+}$--OH groups.
In addition, such a charge has been overall explained by the
theories of binary oxides \cite{Lenglet}. There will be shown effects of the Ps formation under a
positron irradiation of V$_2$O$_5$--H$_2$O sols straight away
after a magnetic field effect (the magnetic splitting
of Ps levels will not be taken into account).

\subsection{Positron annihilation in water. Analogy with aqueous dioxane}
Consider an example of positron and muon studies of water-dioxane mixtures \cite{dioxane}, which revealed the dependence of the $o$-Ps formation on the component rate. This means that there are three regions of relative concentrations $x$ of water in dioxane ($x<0.2$, $0.2<x<0.8$ and $x>0.8$). First region corresponds to the presence of water clusters forming by hydrogen bonds, which are able to trap electrons, in such a cluster, at excess of electrons, a hydrated electron may appear. Negative charged water clusters (H$_2$O)$^-_n$, $n>$2 allow Ps to exist in dioxane phase and retain constant its lifetime. At $0.2<x<0.8$, at the water-dioxane phase segregation, giant water globules \cite{Naberu en}, \cite{Naberu ru} ($n=500-600$) allow Ps to depart to the dioxane phase. Eventually, regimes of the Ps formation differ with clusterization corresponding to different concentrations.

This analogy serves to identification of two types of ordering
in the sol phase of V$_2$O$_5$--H$_2$O, interacting with the positron beam. Mass of the V$_2$O$_5$
molecule is in order large than of the water molecule. V$_2$O$_5$ molecules are joint into
nematic tactoids, and water have to be globular at $n=500$ similarity to the aqueous dioxane aggregation in frames of the representation \cite{dioxane}.

\begin{figure}[h]
\includegraphics[width=16pc]{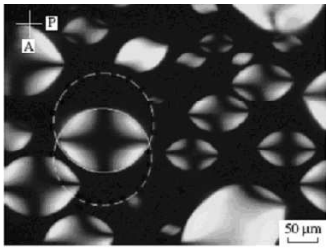}\hspace{2pc}%
\begin{minipage}[b]{18pc}\caption{\label{f1} \small The texture of tactoid phase in a vanadium pentoxide-water system with the concentration of V$_2$O$_5$ of 1.1$\%$ by mass. The cell thickness is 200$\mu$m, \cite{Kazna}.}
\end{minipage}
\end{figure}

The result of e$^+$ interactions with water is well-known \cite{Mogensen}, \cite{Byak Nichip}, \cite{Gold}. Ps forms due to the sequence of reactions, the initial of which proceed  $ <$10$^{-11}$ s and produce the next active forms:
H$_3$O$^+$, hydrated electron e$^-_{aq}$ and OH group. The second stage ($\leq$ 10$^{-9}$ s) contains a creation of molecular products and a trapping of initial products.

Ps lifetimes of pure water are $\tau_1$=0.12 ns, $\tau_2$=0.40 ns, and $\tau_3$=1.80 ns \cite{polar D}.

\subsection{Specific scales of V$_2$O$_5$ particles and Monte Carlo data}
Evaluations received from \cite{Ind} and \cite{Mich} define averaged interatomic distances $d$ and the respective molecular size the (010) layer configuration, $d_{V-O}$=1.89${\AA}$ \cite{Ind} ($1.79 \ldots 2.01{\AA}$ \cite{Mich}) and O--V--O angles, 86.18$^o$, for all bridging oxygen centers. Terminal vanadyl oxygen centers are assumed directly
on top of vanadium centers in the direction normal to the
(010) layer with $d_{V-O}$=1.58${\AA}$. This results in local VO$_5$ pyramids with square base
and fourfold rotation symmetry in addition to the inversion,
twofold rotation, and mirror symmetries of the (010) layers.
This configuration is similar to the ribbon structure \cite{PD}, where each ribbon is
made of two V$_2$O$_5$ planes at a distance of 2.8${\AA}$.
Roughly speaking, colloid particles are made of ribbon-like polymeric
particles about 0.5 $\mu$m in length and 20 nm wide \cite{PD}. Accept that their charge is set by the Bjerrum length, then it equals 2.040816 e$^-$. Then let a tactoid contains approximately10 polimeric ribbons.

Thermodynamics of nematic tactoids should be balanced under electrostatic, van der
Waals and excluded volume interactions.

The excluded volume concept means the Onsager theory of the LC order parameter, widely applicable for nematics.
The Monte Carlo (MC) simulations for LC tactoids were developed \cite{Allen}, \cite{Bates} in respect to the statistical analysis of the order parameter behavior, taking into account a bipolar shape of nematic inorganic droplets and their surface energy causing their formation. In the
lattice-gas Humphries-Luckhurst-Romano model \cite{Bates}, molecules of the inner region of a droplet tend to
be aligned with the major axis of the droplet as the system
is cooled below a tricritical point, verging towards one of the equilibrium states.

\subsection{Ps spectra of V$_2$O$_5$--H$_2$O sols and coalescence of droplets }
The correspondence to the polymer ribbon configuration in semiconducting  V$_2$O$_5$ molecules and possibility of tactoids to keep the inter plane distance (2.8${\AA}$) allow us to apply a number of approaches to calculate spectra of sols.

Applying the model of fluid-filled pores \cite{Bug}, \cite{Dupasquier} as the self-annihilation rate $k\Gamma_0$ and pick-off annihilation rates from fluid $\Gamma_{p.o. fluid}$ and pore walls $\Gamma_{p.o. wall}$, we can consider
the summary annihilation rate as
\begin{equation}
\tau^{-1}\equiv \Gamma=k\Gamma_0+\Gamma_{p.o. fluid}+\Gamma_{p.o. wall}.
\end{equation}
Here
\begin{equation}
\Gamma_{p.o. wall}=(2ns^{-1})\int_{r=R_c-\Delta}^{r=R_c}n_+({\mathbf r})dr^3,
\end{equation}
 $n_{+/-}({\mathbf r})$ is the positron(/electron) density at the direction ${\mathbf r}$ of a tactoid sphere (though here we mean bipolar geometry of a tactoid \cite{Kazna}, see below), in our case $\Delta$=2.8${\AA}$.
\begin{equation}
\Gamma_{p.o. fluid}=\pi r^2_cc\int n_-({\mathbf r})  n_+({\mathbf r})\gamma [n_-({\mathbf r})] dr^3.
\end{equation}

For the fluid component we need to apply standard numerical methods (e.g. \cite{Puska} and so on). In the exact calculations, one may take into account only e$^+$ density, whereas electrons in V$_2$O$_5$ ribbons were counted out in a manner of \cite{PD}.

In the approach independent on the initial positron energy, we rest at the important geometrical properties of tactoids, such as their bispherical geometry leading to polarity and nontriviality in the definition of $R_c$ \cite{Kazna}. Namely, $\Gamma_{p.o. wall}$ is not proportional to the fourth degree of $r$, as it follows from (2).

The coalescence of the V$_2$O$_5$'s tactoids was observed by Kaznacheev \cite{Kazna} in the absent of a magnetic field. In the sol region of V$_2$O$_5,n$H$_2$O, where $n$=250-800, the interparticle distance equals to 400-800${\AA}$ \cite{Sonin}. Whiles volume of two bonded tactoids will be increased in two times, the $\tau_3$ Ps lifetime should decrease in eight times.

Besides of this, estimations of Ps lifetimes of some glasses doped with V$_2$O$_5$ \cite{egypt} shows that depending on the mole V$_2$O$_5$ content, in vicinity of 1 mol $\%$, lifetime slopes and remains constant at the increase of pentoxide fraction in the glass, positrons bond with non-bridging oxyden changing V$^{5+}$ ions onto paramagnetic V$^{4+}$ and V$^{3+}$. So, the mechanism of the Ps $ortho$-$para$ conversion may be considered to explain a positronium states.

\section{Conclusion}
From the above presentation, we may conclude, that apart from the consideration of cooperative effects in water leading to clusterization of negative charges, the coalescence of sol particles of aqueous V2O5 is the effect of a decrease of $\tau_3$ ortho-Ps lifetime achieved regards to condensation of a giant negative charge in bulk tactoids. It is a further measurable effect.

\end{document}